\documentclass[pra,twocolumn,aps,showpacs,superscriptaddress]{revtex4-1}
\usepackage{epsfig,graphicx,tabularx,comment}
\usepackage{psfrag}
\usepackage{graphics}
\usepackage{epsfig}
\usepackage{bm}
\usepackage{verbatim,color,ulem}
\usepackage{hyperref, comment}
\usepackage{amssymb}
\usepackage{physics}
\usepackage{mathrsfs}

\newcommand{\beq}{\begin{equation}}
\newcommand{\eeq}{\end{equation}}
\newcommand{\beqa}{\begin{eqnarray}}
\newcommand{\eeqa}{\end{eqnarray}}
\newcommand{\bal}{\begin{aligned}[b]}
\newcommand{\eal}{\end{aligned}}

\newcommand{\cred}{\color{black}}
\newcommand{\cblue}{\color{black}}

\begin{document}

\title{Dissipation and fluctuations in elongated bosonic Josephson junctions}

\author{F. Binanti}
\affiliation{Dipartimento di Fisica e Astronomia 'Galileo Galilei', 
Universita di Padova, via Marzolo 8, 35131 Padova, Italy}

\author{K. Furutani}
\affiliation{Dipartimento di Fisica e Astronomia 'Galileo Galilei', 
Universita di Padova, via Marzolo 8, 35131 Padova, Italy}
\affiliation{INFN - Sezione di Padova, via Marzolo 8, 35131 Padova, Italy}

\author{L. Salasnich}
\affiliation{Dipartimento di Fisica e Astronomia 'Galileo Galilei', 
Universit\`a di Padova, via Marzolo 8, 35131 Padova, Italy}
\affiliation{INFN - Sezione di Padova, via Marzolo 8, 35131 Padova, Italy}
\affiliation{Padua Quantum Technologies Research Center, 
Universita di Padova, \\ via Gradenigo 6/b, 35131 Padova, Italy}
\affiliation{CNR-INO, via Nello Carrara 1, 50019 Sesto Fiorentino, Italy}

\begin{abstract}
We investigate the dynamics of bosonic atoms in elongated 
Josephson junctions. We find that 
these systems are characterized by an intrinsic coupling between 
the Josephson mode of macroscopic quantum tunneling and the sound modes. 
This coupling of Josephson and sound modes gives rise to a 
damped and stochastic Langevin dynamics for the Josephson 
degree of freedom. From a microscopic Lagrangian, we deduce 
and investigate the damping coefficient and the stochastic noise, 
which includes thermal and quantum fluctuations. 
Finally, we study the time evolution 
of relative-phase and population-imbalance fluctuations of the Josephson 
mode and their oscillating thermalization to equilibrium. 
\end{abstract}


\maketitle

\section{Introduction}

In the last forty years dissipative quantum systems 
under the effect of random noise have been extensively 
investigated both theoretically and experimentally 
\cite{caldeira,ford,schmidt,bulgadaev,dalla}. 
An important tool in the study of dissipative systems is the 
Caldeira-Leggett model \cite{caldeira}, where a bath of harmonic 
oscillators is coupled to the main system. 
This bath leads to damping and noise in the main system and the 
dynamics is governed by the generalized Langevin 
equation \cite{caldeira,ford}. One can find such dynamics in many 
different physical systems, for instance in superconducting Josephson 
circuits \cite{koch80,koch82}.

Recently, the dynamics of ultracold atoms in a double-well potential 
(bosonic Josephson junction) has attracted marked 
attention \cite{minguzzi,pigneur,tononi}. Ref. \cite{minguzzi} has revealed 
the presence of intrinsic coupling between the Josephson mode and 
sound modes leading to damping of oscillations of population 
imbalance in a one-dimensional bosonic Josephson junction. 
The peculiarity of this system is the intrinsic coupling between 
the Josephson mode and the bath which is distinct from a system in contact 
with an external bath like the Caldeira-Leggett model \cite{caldeira}. 
Namely, the one-dimensional Bose Josephson junction exhibits quantum 
Langevin dynamics in the Josephson mode without any external reservoir.
Furthermore, the typical approach of a Caldeira-Leggett model consists
in the introduction of an external bath in a phenomenological way, as well
as for the interaction between the system and the environment. Here instead
we manage to pull out an intrinsic phonon bath (interacting with the Josephson
modes) straightly from the model.

In this paper we start from a model of two one-dimensional quasi-condensates 
with a Josephson coupling in the head-to-tail configuration discussed in 
Refs. \cite{minguzzi,tononi}. 
We treat phase and number density as classical fields to obtain 
the effective Lagrangian of relative phase.
By adopting this different procedure with respect 
to Ref. \cite{minguzzi}, we show that this system involves the 
intrinsic coupling between the Josephson mode and other infinite numbers 
of sound modes. Consequently, the dynamics of the Josephson mode 
is described by the quantum Langevin equation with damping and noise 
that includes thermal and quantum fluctuations. 
We find an analytic formula for the damping coefficient 
moreover, we derive analytic expressions for the 
fluctuations of the Josephson relative phase and Josephson 
population imbalance in the high-temperature regime, 
extending the results of Ref. \cite{minguzzi}. 
Finally, we analyze numerically the fluctuations of the Josephson mode 
in the general case, in particular in the low-temperature regime 
where quantum fluctuations play a crucial role. Remarkably, 
in this low-temperature regime, an ultraviolet cutoff for the sound 
modes is needed. In this way, we determine the thermalization 
to the equilibrium of the Josephson fluctuations.   

\section{Bosons with Josephson tunnelling}

We start from the following Lagrangian density which consists of 
two weakly-interacting Bose-Einstein quasi-condensates ($j=1,2$) 
made of atoms with mass $m$ in one dimension
\beq
\bal
\mathscr{L}
& = \sum_{j=1}^{2}\left[ i\hbar \psi_j^* \partial_t \psi_j 
-  \frac{\hbar^2}{2m}\abs{\partial_x \psi_j}^2 - 
\frac{g}{2}\abs{\psi_j}^4 \right] \\
&+ \frac{J(x)}{2}\left[\psi_1^* \psi_2 + \psi_2^* \psi_1\right] ,
\eal
\label{lagden}
\eeq
where $\hbar$ is the Planck constant. Here $g$ is the strength 
of the inter-atomic potential 
in the contact-interaction approximation and $J(x)$ is the 
space dependent tunneling-energy coupling. 
The complex field $\psi_j(x,t)$ of the $j$-th quasi-condensate 
can be rewritten by means of the Madelung representation
\beq
\psi_j(x,t) = \sqrt{\rho_j(x,t)} \, e^{i\phi_j(x,t)} ,
\label{madelung}
\eeq
where $\rho_j(x,t)=|\psi_j(x,t)|^2$ is its atomic density. 
Substituting the expression of Eq. (\ref{madelung}) into the 
Lagrangian in Eq. (\ref{lagden}), we obtain
\begin{eqnarray}
\mathscr{L} &=& \sum_{j=1}^{2}\Bigg[\frac{i\hbar}{2}\partial_t\rho_j 
- \hbar\rho_j\partial_t\phi_j - \frac{\hbar^2}{2m}\Bigg[ \frac{1}{4\rho_j}
(\partial_x \rho_j)^2 
\nonumber  
\\
&+& \rho_j(\partial_x \phi_j)^2 \Bigg] - 
\frac{g}{2}\rho_j^2 \Bigg] + J(x)\sqrt{\rho_1\rho_2}\cos(\phi_1 - \phi_2) .
\label{lagmadelung}
\end{eqnarray}
A compact description of the system is reached when we introduce the 
relative phase and the population imbalance 
\begin{eqnarray}
\phi &=& \phi_1 - \phi_2 ,
\\
\zeta &=& \frac{\rho_1 - \rho_2}{2\bar\rho} ,
\end{eqnarray}
with $\bar\rho = (\rho_1 + \rho_2)/2$ the average atomic density. 
Introducing also a total phase $\bar\phi = \phi_1 + \phi_2$,
one can express the Lagrangian in Eq. (\ref{lagmadelung}) 
in terms of these quantities. In addition, let us work in a canonical ensemble 
(no particle leaves or enters the system) for which 
$\partial_t\bar\rho = 0$ holds, and assuming it also for the total 
phase $\partial_t\bar\phi = 0$, we obtain a new Lagrangian density 
\begin{eqnarray}
\mathscr{L} &=& -\hbar\bar\rho\zeta\dot\phi - 
\frac{\hbar^2\bar\rho}{4m} (\partial_x\phi)^2 - 
g(\bar\rho^2 + \bar\rho^2\zeta^2) 
\nonumber 
\\
&+& J(x)\bar\rho\sqrt{1-\zeta^2}\cos(\phi) ,
\end{eqnarray}
in which we also assumed space homogeneity 
$\partial_x \bar\rho = \partial_x \bar\phi = 0$ and neglected space 
variations of the population imbalance $\partial_x \zeta = 0$ 
{\cred (see also, for instance, Refs. \cite{tononi,essler})}. 
It is reasonable (also with the experimental setup) to work with small values
of the population imbalance, such that $\abs{\zeta} \ll 1$. 
In this case we obtain the equation of motion 
\beq
\zeta(x,t) = -\frac{\hbar\dot\phi(x,t)}{2g\bar\rho + J(x)\cos(\phi)} .
\label{zetaeom}
\eeq
Setting the Josephson regime $2g\bar\rho \gg J(x)$, we can simplify 
the expression of Eq. (\ref{zetaeom}) and the Lagrangian density only for
the relative phase can be written as 
\beq
\mathscr{L} = \frac{\hbar}{4g}\dot\phi^2 - 
\frac{\hbar^2\bar\rho}{4m}(\partial_x\phi)^2 + J(x)\bar\rho\cos(\phi) .
\label{lagdenphi}
\eeq

\begin{figure}[t]
\centerline{\epsfig{file=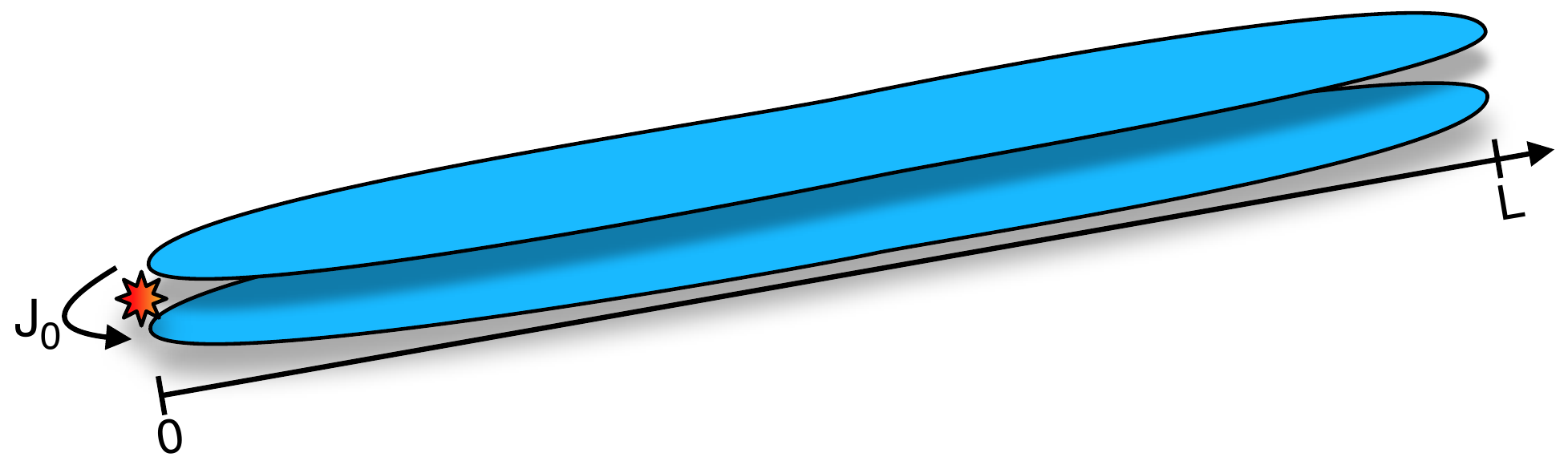,width=8cm}}
\small 
\caption{Head-to-tail configuration of the system under investigation. 
The tunneling of bosons occurs with {\cred strength} $J_0$ only at $x=0$. 
{\cred Here $J_0$ has the dimensions of energy times length}.} 
\label{fig1}
\end{figure}

{\cred In the rest of the paper, we shall adopt the 
head-to-tail configuration,  as depicted in Fig. \ref{fig1}, where 
\beq
J(x) = J_0 \, \delta(x) \; . 
\label{ddf}
\eeq 
with $J_0$ a constant tunneling coupling, $x\in [0,L]$ and $L$ 
the length of the two elongated Bose-Einstein quasi-condensates.} 
This head-to-tail configuration is equivalent 
to a system where one quasi-condensate is 
confined in the region $[-L,0]$ while the other quasi-condensate 
is confined in the region $[0,L]$, 
{\cred and the tunneling barrier is located at $x=0$. 
This is the same configuration considered in Ref. \cite{minguzzi}. 
As we shall see, the Dirac delta function gives rise to the 
so-called boundary sine-Gordon model \cite{fendley}. 
More generally, the idea of approximating a realistic 
finite-range tunneling energy $J(x)$ with a zero-range one 
$J_0\delta(x)$ implies that $J_0=\int_{-L}^{L} J(x) dx={\tilde J}(k=0)$, 
where ${\tilde J}(k)$ is the Fourier transform of $J(x)$.
This approximation is justified if the 
characteristic range of $J(x)$, localized around $x=0$,  
is much smaller than the characteristic 
lengths of the problem under consideration.} 
{\cblue Within this approximation one can obtain the tunneling coupling 
given by Eq. (\ref{ddf}) starting from the Lagrangian density 
of a single bosonic field $\psi(x,t)$, with $x\in [-L,L]$, setting 
$\psi(x,t)=\psi_1(x,t)+\psi_2(x,t)$ where $\psi_1(x,t)$ is mainly 
localized on one side of a narrow potential barrier $U(x)$ with maximum  
at $x=0$ and $\psi_2(x,t)$ is mainly localized on the other side of $U(x)$. 
In this case one gets $U(x) |\psi(x,t)|^2\simeq -J_0\delta(x) 
[\psi_1^*(x,t)\psi_2(x,t)+\psi_1(x,t)\psi_2^*(x,t)]$, with 
$U(x)\simeq -J_0\delta(x)$. Alternatively, the delta-function tunneling 
term of Eq. (\ref{ddf}) can be intepreted as due to a local internal 
Rabi coupling. Experimentally one can induce a Rabi coupling 
in a small region of a Bose-Bose mixture.}

{\cred It is important to stress that the case of a uniform tunneling 
coupling, i.e. $J(x)=J_{\mathrm{unif}}$, which is relevant for the experiment of 
Pigneur {\it et al.} \cite{pigneur}, was theoretically considered by 
Bouchoule \cite{bouchoule} and also by Grisins and Mazets \cite{grisins}.} 

\section{Quasi-particle description}

We will now introduce a quasi-particle description for the phase, 
based on the following mode expansion 
\beq
\phi(x,t) = \frac{1}{\sqrt{L}}\sum_{n=0}^{+\infty}q_n(t)\Phi_n(x) ,
\label{phiexpansion}
\eeq
where $q_n(t)$ are coordinates, and $\Phi(x)$ are real eigenfunctions 
satisfying $-\hbar^{2}/\left(2m\right)\partial_{x}^{2}\Phi_{n}(x)
=\epsilon_{n}\Phi_{n}(x)$ where $\epsilon_{n}=\hbar^{2}k_{n}^{2}/\left(2m\right)$ 
and $k_{n}=\pi n/L$, and constituting an orthonormal 
basis $\int_{0}^{L} \Phi_n(x)\Phi_m(x) \, dx = \delta_{n,m}$. 
This mode expansion in Eq. \eqref{phiexpansion} leads to
\begin{eqnarray}
{\cal L} = \int_{0}^{L} \mathscr{L} \, dx &=& \frac{M}{2}\sum_n\dot q_n^2  
-\frac{M}{2}\sum_n\omega_n^2 q_n^2 
\nonumber
\\
&+& J_0 \bar\rho \cos{\left(\frac{1}{L}\sum_n q_n\right)} .
\end{eqnarray}
Here we have defined the effective mass 
\beq
M=\frac{\hbar^2}{2gL} ,
\eeq
and the mode dispersion
\beq
\omega_n = c_s \, k_n ,
\eeq
where 
\beq 
c_s = \sqrt{\frac{\bar\rho g}{m}} ,
\label{imp-sound}
\eeq
is the speed of sound, as known from the description of 
superfluids, and that $J_0$ and $g$ have the dimension 
of energy times a unit of length.

We can distinguish between the Josephson mode 
we are interested in $(n=0)$ and a phonon bath
(we mention phonons due to the acoustic spectrum above) made of infinite 
contributions, that affect the evolution of the oscillations
in the Josephson mode. 
Separating the Josephson mode from bath modes, one obtains
\beqa
{\cal L} &=& \frac{M}{2}\dot q_0(t)^2 + \frac{M}{2}\sum_{n=1}^{+\infty}
\dot q_n(t)^2 - \frac{M}{2}\sum_{n=1}^{+\infty}\omega_n^2 q_n(t)^2 
\nonumber 
\\
&+& J_0\bar\rho\cos{\left[\frac{q_0(t)}{L} + 
\frac{1}{L}\sum_{n=1}^{+\infty}q_n(t)\right]} .
\label{qlagrangian}
\eeqa
This is the Lagrangian describing the Josephson mode coupled 
to a bath composed of independent harmonic oscillators. 

\section{Damped dynamics}

The Legendre transformation $\mathcal{H} = 
\sum_{n=0}^{+\infty}\dot q_n(t)p_n(t) - {\cal L}$ with $p_n(t) = M\dot q_n(t)$ 
results in the following Hamiltonian 
\beq
\mathcal{H} = \sum_{n=0}^{+\infty}\left[ \frac{p_n^2}{2M} + 
\frac{M\omega_n^2}{2}q_n^2 \right] - J_0\bar\rho\cos(\frac{1}{L}
\sum_{n=0}^{+\infty}q_n(t)) .
\eeq
Now we perform a canonical transformation with 
new coordinates and momenta 
\begin{align}
& Q_0(t) = q_0(t) + \sum_{n=1}^{+\infty}q_n(t) , \label{cantransf1} \\
& Q_n(t) = q_n(t) \hspace{14.5mm}\mbox{for}\hspace{3mm} n\neq0 ,
\label{cantransf2} \\
& P_0(t) = p_0(t) , \label{cantransf3} \\
& P_n(t) = p_n(t) - p_0(t) \hspace{3mm}\mbox{for}\hspace{3mm} 
n\neq0 . \label{cantransf4}
\end{align}
Notice that is this is the analogue of the unitary transformation 
used in Ref. \cite{minguzzi} for the quantum operators. 
In this way, we obtain the transformed Hamiltonian 
\begin{equation}
\mathcal{H} = \frac{P_0^2}{2M} + \sum_{n=1}^{+\infty}
\left[ \frac{(P_0 + P_n)^2}{2M} + \frac{1}{2}M\omega_n^2Q_n^2 \right] 
- J_0\bar\rho\cos(\frac{Q_0}{L}) .
\label{transfhamiltonian}
\end{equation}
It is worth noting that the canonical transformation 
introduced an intrinsic coupling
in the harmonic oscillators of the bath, between the zero-mode and 
the excited modes.
{\cred The Hamiltonian of Eq. \eqref{transfhamiltonian} corresponds 
to the velocity-coupling model in which the coupling is through 
the momentum \cite{ford}. For the sake of completeness, we stress 
that in the case of two parallel tubes with uniform tunneling energy, 
i.e. with $J(x)=J_{\mathrm{unif}}$, only the inclusion of anharmonic terms 
gives rise to a coupling between the Josephson mode 
and the bath of elementary excitations.}

Hamilton equations provided by the Hamiltonian 
in Eq. (\ref{transfhamiltonian}) are
\begin{align}
& \dot Q_0(t) =  \frac{P_0(t)}{M} + \sum_{n=1}^{+\infty}\frac{P_0(t)+P_n(t)}{M} ,
\label{dotQ0}\\
& \dot Q_n(t) = \frac{P_0(t) + P_n(t)}{M} ,\label{dotQn}\\
& \dot P_0(t) = -\frac{J_0\bar\rho}{L}\sin(\frac{Q_0(t)}{L}) ,\label{dotP0} \\
& \dot P_n(t) = -M\omega_n^2Q_n(t) .\label{dotPn}
\end{align}
These equations of motion lead to
\beqa
Q_n(t) &=& \cos(\omega_n t)Q_n(0) + \frac{\sin(\omega_n t)}{\omega_n} 
\dot Q_n(0) 
\nonumber 
\\
&-& \frac{J_0\bar\rho}{ML\omega_n^2}
\biggl[\sin(\frac{Q_0(t)}{L}) - \cos(\omega_n t)\sin(\frac{Q_0(0)}{L}) 
\nonumber 
\\
&-& \int_{0}^{t}dt' \cos{\left[\omega_n(t-t')\right]}\cos(\frac{Q_0(t')}{L})
\frac{\dot Q_0(t')}{L} \biggl] .
\label{Qnt}
\eeqa 
Details for the derivation of Eq. \eqref{Qnt} are summarized 
in Appendix \ref{AppA}. 
One can easily show that 
the Josephson mode is related to the relative phase at $x=0$ as 
\beq
\phi_0(t) = \phi(x=0,t) = \frac{Q_0(t)}{L} .
\eeq
Eventually, we reach the equation of motion for a damped harmonic 
oscillator with respect to the relative phase
\beq
\ddot \phi_0(t) + \int_{0}^{t}dt' \gamma(t-t')\dot\phi_0(t') + 
\Omega_0^2\sin(\phi_0(t)) = \xi_\phi(t) ,
\label{eqdampedphi}
\eeq
in which the damping is ruled by the damping kernel
\beq
\gamma(t-t') = \Omega_0^2\sum_{n=1}^{+\infty}\cos(\omega_n(t-t'))
\cos(\phi_0(t')) ,
\label{gamma}
\eeq
proportional to the Josephson frequency defined as 
\beq
\Omega_0 = \sqrt{\frac{\bar\rho J_0}{ML^2}} .
\label{imp-fre}
\eeq
Now let us focus on the case of a small relative phase 
$\abs{\phi_0(t)}\ll 1$. In this case $\cos(\phi_0(t'))\sim 1$ holds.
The expression of Eq. (\ref{gamma}) can be calculated explicitly by 
moving to the continuum limit as
\beq
\gamma(t-t') = \gamma_0 \, \delta(t-t') ,
\eeq
where 
\beq 
\gamma_0 = \frac{\bar\rho J_0}{MLc_s} ,
\label{imp-gamma}
\eeq
is the damping constant that rules the relaxation 
in time of the phase. Using the damping constant, 
one can write Eq. \eqref{eqdampedphi} as 
\beq
\ddot \phi_0(t) + \gamma_0\dot\phi_0(t) + \Omega_0^2\phi_0(t) = 
\xi_\phi(t) .
\label{eqdampedphi2}
\eeq
This is the main equation in the paper. On the left-hand side of this equation, 
it involves the Josephson frequency $\Omega_0$, given by Eq. (\ref{imp-fre}), 
and the damping constant, given by Eq. (\ref{imp-gamma}) with 
the speed of sound $c_s$ of Eq. (\ref{imp-sound}).  

The right-hand side of Eq. (\ref{eqdampedphi2}) is a noise term 
denoted by $\xi_\phi(t)$, composed of the infinite contributions of the 
bath modes and dependent only on the initial conditions of coordinates 
and momenta 
\beq
\begin{split}
\xi_\phi(t) = &-\sum_{n=1}^{+\infty}
\left[ \omega_n^2\cos(\omega_n t)\frac{Q_n(0)}{L}  
+ \omega_n\sin(\omega_n t)\frac{\dot Q_n(0)}{L} \right] \\
& - \gamma_0\delta(t)\sin{\left(\phi_0(0)\right)} .
\label{xiphi0}
\end{split}
\eeq
The last term has the nature of a transient, and in order to avoid 
its complications, we will get rid of the transient term by setting the initial 
condition $\phi_0(0) = 0$. 

{\cred \subsection{Deterministic dynamics of the Josephson mode}

As an homogeneous solution of Eq. (\ref{eqdampedphi2}) without the noise, 
one gets}
\beq
\phi_0(t) = e^{-\gamma_Q \Omega_0 t}\frac{\dot \phi(0)}{\Omega_0\gamma_J}
\sin(\gamma_J \Omega_0 t) ,
\label{phihomogeneous}
\eeq
where we defined 
\beq 
\gamma_Q = {\gamma_0\over 2\Omega_0} ,
\label{luca-gammaq}
\eeq
and 
\beq 
\gamma_J = \sqrt{1-\gamma_Q^2}  .
\label{luca-gammaj}
\eeq
We can confirm that the dimensionless damping constant $\gamma_{Q}$ 
corresponds to $\sqrt{E_{J}/K}$ in Ref. \cite{minguzzi} in the 
weak-coupling limit where $K\sim1/\sqrt{g}$ denotes the Luttinger 
parameter in Luttinger liquid theory \cite{minguzzi}. This indicates 
that our current analysis under the quasi-particle description recovers 
the analysis within Luttinger liquid theory in the weak-coupling 
limit $K\sim1/\sqrt{g}$.
In the following, only the underdamped regime $\gamma_{Q}<1$ is 
analyzed for simplicity. It reflects the most similar behaviour with 
respect to the experimental
observations of the relative phase and population imbalance \cite{pigneur}.
The other two regimes, overdamped and critical damped, are 
respectively obtained in the case of $\gamma_Q > 1$ and $\gamma_Q = 1$.

\begin{figure}[t]
\centerline{\epsfig{file=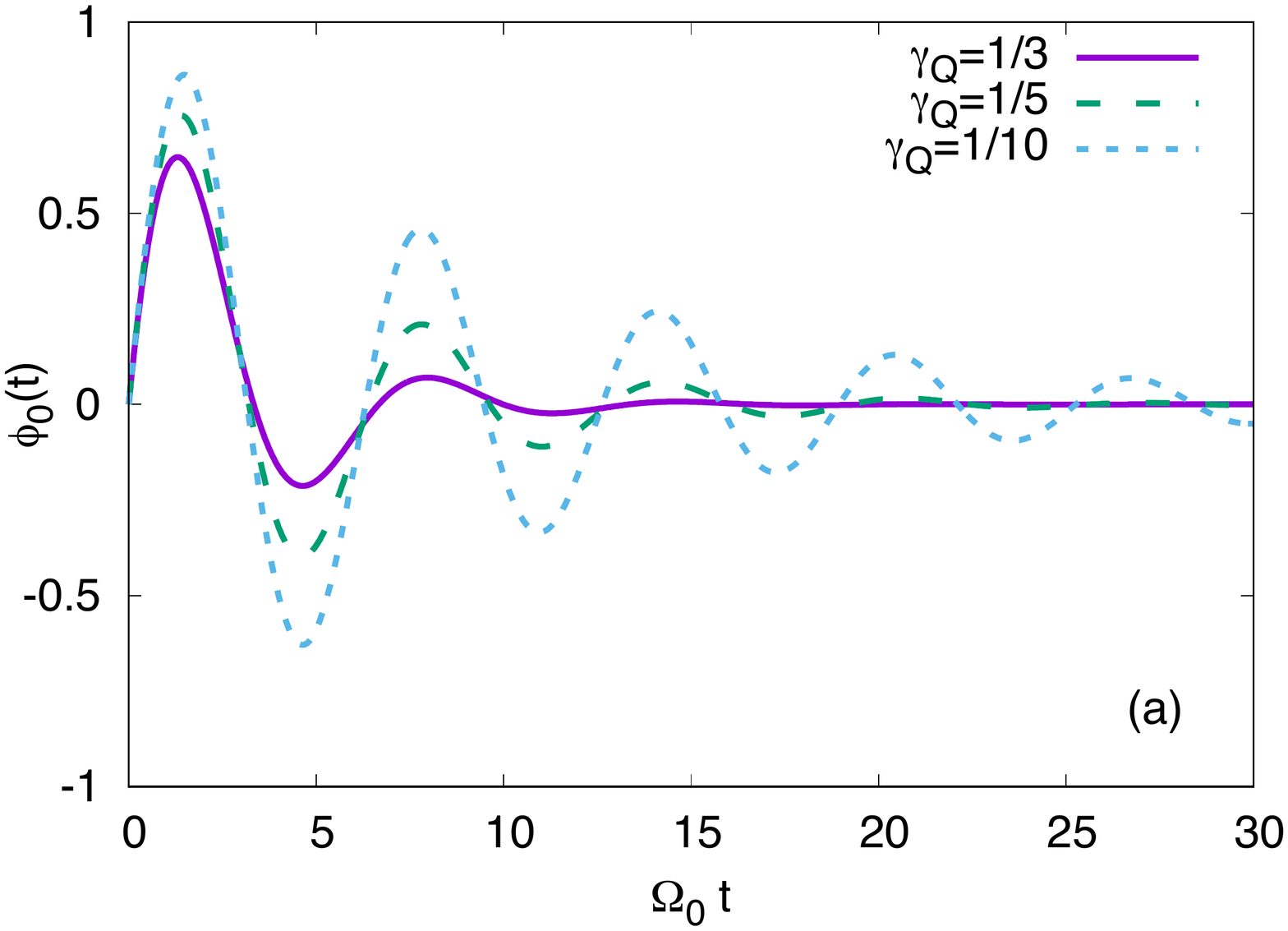,width=1\linewidth}}
\centerline{\epsfig{file=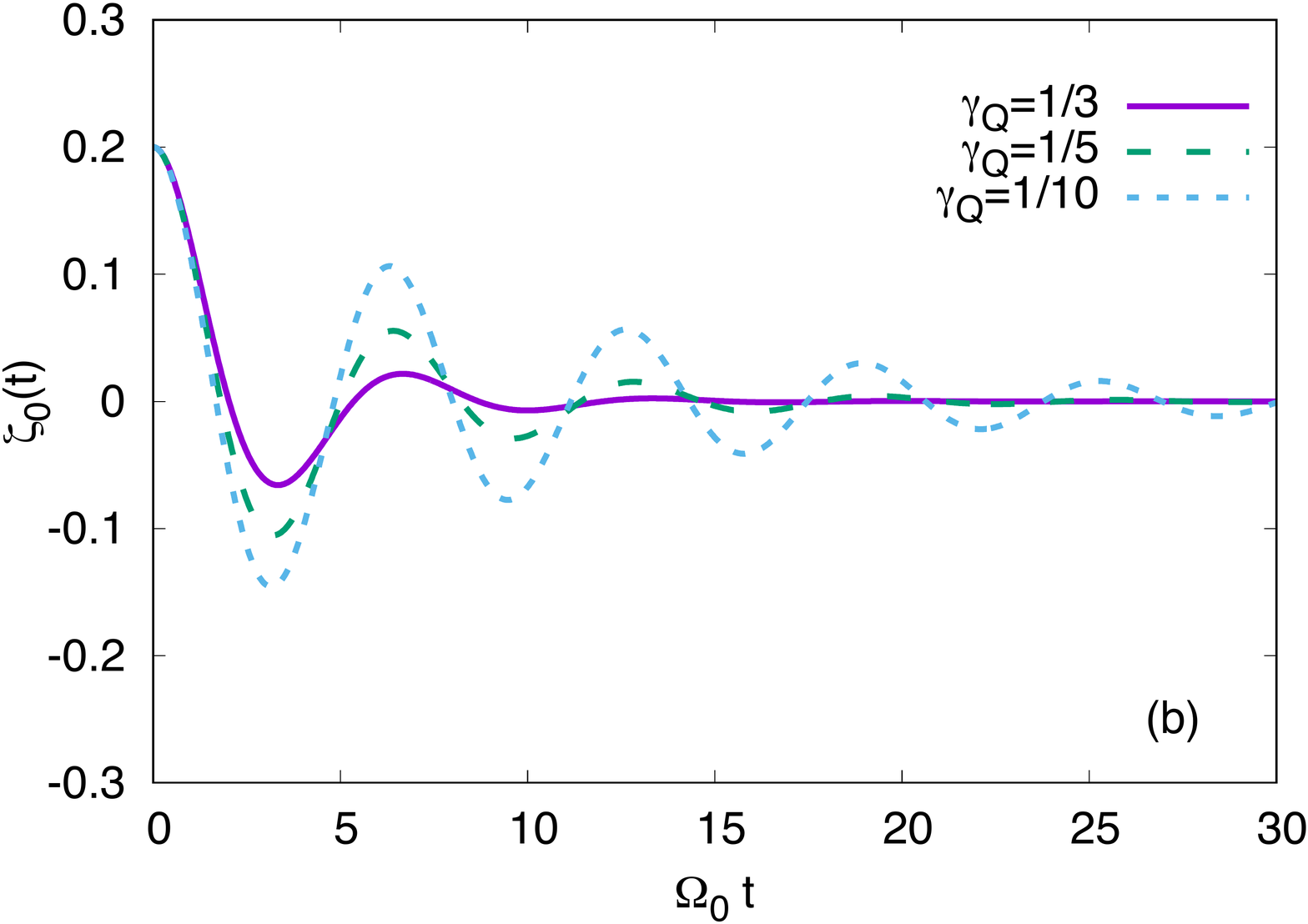,width=1\linewidth}}
\small 
\caption{Josephson mode dynamics without the effect of the noise 
($\xi_\phi(t) = 0$). Time evolution of the relative phase $\phi(t)$ (panel (a)) 
and of the the zero-mode $\zeta_0(t)$ of the population imbalance 
(panel (b)). Here we consider the underdamped case with $\gamma_{Q}<1$.  
Initial conditions $\phi_0(0) = 0$ and $\dot \phi_0(0)/\Omega_0 = 1$; 
$\zeta_0(0) = 0.2$ and $\dot \zeta_0(0)/\Omega_0 = 0$.} 
\label{fig2}
\end{figure} 

In panel (a) of Fig. \ref{fig2}, we plot the dynamics of the relative 
phase $\phi_0(t)$, without the effect of noise $\xi_\phi(t)$ 
and setting the initial condition  $\dot \phi_0(0)/\Omega_0 = 1$ 
for simplicity. In panel (b) of Fig. \ref{fig2}, we plot instead 
the corresponding population imbalance $\zeta_0(t)$, with 
initial conditions $\zeta_0(0) = 0.2$ and $\dot \zeta_0(0)/\Omega_0 = 0$.
The initial conditions for $\dot\zeta(0)/\Omega_0$ is dependent on the choice
of $\phi_0(0)$ through Eq. (\ref{dotP0}). On the other hand, the choice
of $\zeta_0(0)$ is necessary in order to start from a configuration out of
equilibrium, in which the number of atoms in the two wells is not balanced. 
It illustrates quite intuitive behaviours. As one increases the damping 
coefficient $\gamma_{0}$, the amplitude od oscillation is suppressed 
and both of $\phi_{0}(t)$ and $\zeta_{0}(t)$ vanish for $t\to\infty$.

The results in Fig. \ref{fig2} is, however, missing the relevant role of the 
noise, which produces fluctuations around the 
mean-field behaviour shown in the figure. In general, 
these fluctuations have both quantum and thermal components. 
We investigate these effects of the noise in the next Sections. 

{\cred\subsection{Dynamics of the Josephson mode in the presence
of noise}}

We refer to $\xi_\phi(t)$ as a quasi-stochastic noise due to the analogy 
of Eq. (\ref{eqdampedphi2}) with a generalized Langevin equation.
The stochastic nature of this term, however, lies only in 
the initial conditions $Q_n(0)$ and $\dot Q_n(0)$, then $\xi_\phi(t)$ 
is not a random quantity in time.
Thanks to this property, we do not need special tools like stochastic 
calculus in order to integrate Eq. (\ref{eqdampedphi2}) in time, but 
we will find the particular solution using the method of variation 
of parameters. The solution takes the form of 
\begin{equation}
\phi_0(t) = e^{-\gamma_{Q}\Omega_0 t}\frac{\sin(\gamma_{J}\Omega_0 t)}{\gamma_J} 
+ \int_{0}^{t}dt' \chi(t-t') \xi_\phi(t'),
\label{generalsolphi}
\end{equation}
where we have defined the following retarded function 
\beq
\chi(t-t') = \frac{2}{\omega_D}e^{-\gamma_Q\Omega_0(t-t')}
\sin{\left[\frac{\omega_D}{2}(t-t')\right]} \theta(t-t') ,
\eeq
with $\theta(t)$ the Heaviside step function and the oscillation frequency 
\beq 
\omega_D = \sqrt{4\Omega_0^2 - \gamma_0^2} .
\eeq

\section{The population imbalance}

Using Eq. (\ref{zetaeom}) in the Josephson regime, one can make a 
quasi-particle description also 
for the population imbalance $\zeta(x,t)$ as
\beq
\bal
\zeta(x,t) &= \frac{\sqrt{L}}{\hbar\bar\rho}
\sum_{n=0}^{+\infty}p_n(t)\Phi_n(x) \\
&= \frac{P_0(t)}{\hbar\bar\rho} + 
\frac{\sqrt{L}}{\hbar\bar\rho}\sum_{n=1}^{+\infty}\left[P_0(t) + 
P_n(t)\right]\Phi_n(x) .
\eal
\label{zetaexpansion}
\eeq
We can identify the Josephson mode as the first term
\beq
\zeta_0(t) = \frac{P_0(t)}{\hbar\bar\rho}.
\eeq
Equations of motion for the momenta $P_0(t)$ and $P_n(t)$ read
\begin{align}
& \ddot P_0(t) + \Omega_0^2P_0(t) = \Omega_0^2\sum_{n=1}^{+\infty}
\frac{\ddot P_n(t)}{\omega_n^2} \label{P0eom} ,\\
& \ddot P_n(t) + \omega_n^2P_n(t) = -\omega_n^2P_0(t) .\label{Pneom}
\end{align}
They give the damped behaviour of the Josephson mode
\beq
\ddot \zeta_0(t) + \gamma_0\dot \zeta_0(t) + 
\Omega_0^2\zeta_0(t) = \xi_\zeta(t) .
\label{eqdampedzeta}
\eeq
Here we have defined the noise term for the population imbalance as
\begin{equation}
\begin{split}
\xi_\zeta(t) = -\frac{\Omega_0^2}{\hbar\bar\rho}\sum_{n=1}^{+\infty}
\left[ \cos(\omega_n t)P_n(0) + \frac{\sin(\omega_n t)}
{\omega_n}\dot P_n(0) \right] ,
\label{xizeta0}
\end{split}
\end{equation}
where we have set the initial condition $P_0(0) = 0$. 
The meaning of this choice will be stressed in the next section.
The time evolution of $\zeta_0(t)$, without the effect of 
the noise $\xi_\zeta(t)$, is displayed in Fig. \ref{fig2}(b).

\section{Quantum and thermal properties of the noise}

Interesting quantities regarding the noise, which allow us to 
investigate the quantum and thermal fluctuations of $\phi_0(t)$ due to the 
presence of $\xi_\phi(t)$, are its bath average $\expval{\xi_\phi(t)}$ and, 
in particular, its correlation function $\expval{\xi_\phi(t)\xi_\phi(t')}$. 
We will perform a calculation already done in Ref. \cite{ingold} in the context 
of a Caldeira-Leggett model \cite{caldeira}.
In order to evaluate the average over the environment, 
we have to identify the bath Hamiltonian. From Eq. \eqref{transfhamiltonian}, 
denoting the bath part by $\mathcal{H}_{\rm B}$, we have 
\beq
\mathcal{H}_{\rm B} = \sum_{n=1}^{+\infty}\left[ \frac{(P_0 + P_n)^2}{2M} 
+ \frac{M\omega_n^2}{2}Q_n^2 \right] .
\eeq
As we previously pointed out, this Hamiltonian is composed of infinite 
harmonic oscillators which are intrinsically coupled to the system. However, 
it is reasonable to assume that the size of the phonon bath is huge such that 
the system does not affect it. For this reason one can impose $P_0(0) = 0$, 
namely the system is completely decoupled 
from the bath at $t=0$. Hence, in this model, we deal with a bath of 
independent harmonic oscillators \cite{ford} at initial time $t=0$ as
\beq
\mathcal{H}_{\rm B} = \sum_{n=1}^{+\infty}\left[ \frac{P_n^2}{2M} + 
\frac{M\omega_n^2}{2}Q_n^2 \right]
\label{bathhamiltonian}
\eeq
which is used to evaluate the ensemble average. 

In order to evaluate ensemble averages, we consider $Q_n(0)$ and $P_n(0)$ 
to be quantum 
operators. By means of the Hamiltonian in Eq. (\ref{bathhamiltonian}), 
we can adopt the annihilation and creation operators $a_n$ and $a_n^\dagger$ as
\begin{align}
& Q_n(0) = \sqrt{\frac{\hbar}{2M\omega_n}}(a_n + a_n^{\dagger}) ,\\
& P_n(0) = -i\sqrt{\frac{\hbar M \omega_n}{2}}(a_n - a_n^{\dagger}) ,
\end{align}
and then evaluate the average, in the case of $\xi_\phi(t)$ for instance, as 
\beq
\expval{\xi_\phi(t)} = \frac{\Tr\left[e^{-\beta \mathcal{H}_{\rm B}}
\xi_\phi(t)\right]}
{\Tr\left[e^{-\beta \mathcal{H}_{\rm B}}\right]},
\eeq
where $\beta=1/(k_{\rm B} T)$ with $k_{\rm B}$ the Boltzmann constant and $T$ the 
absolute temperature of the bath of oscillators. 

One readily obtains the vanishing averages for both of Eqs. (\ref{xiphi0}) and 
(\ref{xizeta0}) 
\beq
\expval{\xi_\phi(t)} = \expval{\xi_\zeta(t)} = 0 ,
\eeq
while we find the two-point correlation functions as
\beqa
\expval{\xi_\phi(t)\xi_\phi(t')} &=& \sum_{n=1}^{+\infty} 
\frac{\hbar\omega_n^3}
{2ML^2}\Bigg[ \coth(\frac{\beta\hbar\omega_n}{2})
\cos{\left[\omega_n(t-t')\right]} 
\nonumber 
\\
&-& i\sin{\left[\omega_n(t-t')\right]} \Bigg] ,
\label{xiphicorr} 
\eeqa
\beqa
\expval{\xi_\zeta(t)\xi_\zeta(t')} &=& \frac{M\Omega_0^4}{\hbar^2\bar\rho^2}
\sum_{n=1}^{+\infty}\frac{\hbar\omega_n}{2}\Bigg[ 
\coth(\frac{\beta\hbar\omega_n}{2})\cos{\left[\omega_n(t-t')\right]} 
\nonumber 
\\
&-& i\sin{\left[\omega_n(t-t')\right]} \Bigg] .
\label{xizetacorr}
\eeqa
In the high-temperature limit $k_{\rm B} T \gg \hbar\omega_n$, 
one can approximate  
\beq
\coth(\frac{\beta\hbar\omega_n}{2}) \simeq
\frac{2}{\beta\hbar\omega_n},
\eeq
resulting in the following correlation functions 
\begin{align}
& \expval{\xi_\phi(t)\xi_\phi(t')} = -\frac{\gamma_0}{M\Omega_0^2}
k_{\rm B}T\frac{d^2}{dt^2}\delta(t-t') ,\label{hightempxiphicorr} \\
& \expval{\xi_\zeta(t)\xi_\zeta(t')} = \frac{\gamma_{0} M\Omega_0^2}
{\hbar^2\bar\rho^2}k_{\rm B}T\delta(t-t') .\label{hightempxizetacorr}
\end{align}
It is worth noting the fact that $\xi_\zeta(t)$ is a delta-correlated noise 
at high temperature likewise the classical fluctuation-dissipation relation, 
while $\xi_\phi(t)$ is slightly different from a white noise.

In the case of a generic temperature, one has to take into 
account that Eqs. (\ref{xiphicorr}) and (\ref{xizetacorr}) are complex valued 
quantities. In particular it is convenient to separate the real part 
from the purely imaginary one, by rewriting the correlators as 
\beq
\expval{\xi_\phi(t)\xi_\phi(t')} = \frac{1}{2}\expval{\acomm{\xi_\phi(t)}
{\xi_\phi(t')}} + \frac{1}{2}\expval{\comm{\xi_\phi(t)}{\xi_\phi(t')}} .
\label{xicomplexexpansion}
\eeq
The first term, which involves the anti-commutator $\acomm{\cdot}{\cdot}$ 
is related to the real part of Eq. (\ref{xiphicorr}), 
while the second one, which involves the commutator $\comm{\cdot}{\cdot}$, 
describes the purely imaginary term of the noise. 
The anti-symmetric parts are, however, found to result in no contribution 
to the variance of phase or population imbalance {\cred because  
they are odd functions of $\omega$}. 
The Fourier transformed correlators of the symmetric part are given by
\begin{align}
& \int_{-\infty}^{+\infty}dt\expval{\acomm{\xi_\phi(t)}
{\xi_\phi(0)}}e^{-i\omega t} = \frac{2\gamma_0\hbar\omega^3}{ML^2\Omega_0^2}
\coth(\frac{\beta\hbar\omega}{2}) , \label{FTsymmxiphi} \\
& \int_{-\infty}^{+\infty}dt\expval{\acomm{\xi_\zeta(t)}{\xi_\zeta(0)}}
e^{-i\omega t} = \frac{2M\Omega_0^2\gamma_0\hbar\omega}{\hbar^2\bar\rho^2}
\coth(\frac{\beta\hbar\omega}{2}) .\label{FTsymmxizeta}
\end{align}

\section{Fluctuations}

In order to understand the effect of the noise in the dynamics of $\phi_0(t)$ 
and $\zeta_0(t)$, we focus on the calculation of the variance 
$\Delta\phi_0(t)^2=\expval{\phi_0(t)^2} - \expval{\phi_0(t)}^2$ and 
$\Delta\zeta_0^2= \expval{\zeta_0(t)^2} - \expval{\zeta_0(t)}^2$. 
It is obvious from Eq. (\ref{generalsolphi}) that they will depend only on the 
correlators of the noise.

\begin{figure}[t]
\centerline{\epsfig{file=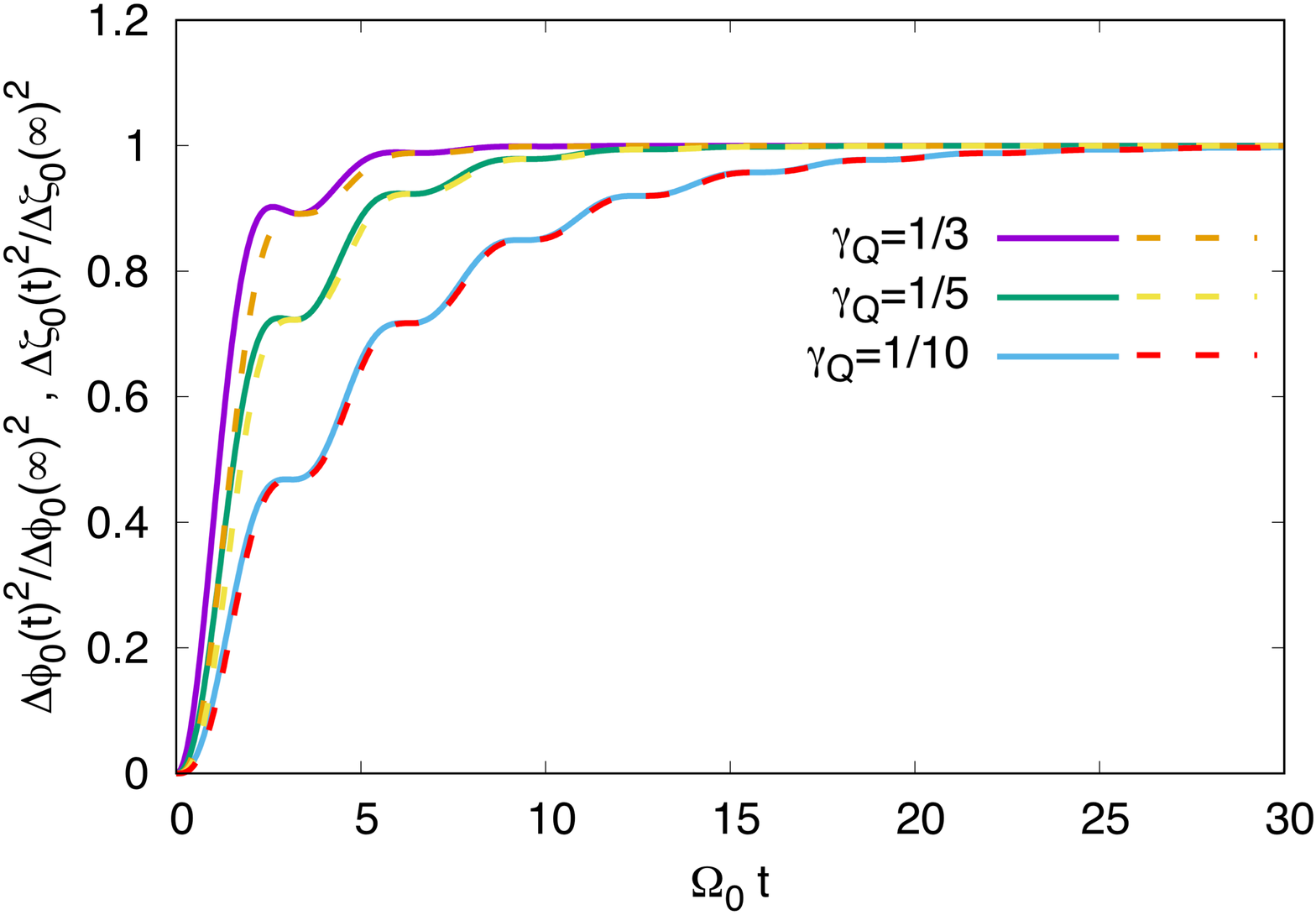,width=1\linewidth}}
\small 
\caption{Relative phase variance $\Delta\phi_0(t)^2$ 
(solid curves) and population imbalance variance $\Delta\zeta_0(t)^2$ 
(dashed curves) of the Josephson mode as a function of time $t$. 
The curves correspond to three different underdamped regimes 
at high temperature, where $k_{\rm B}T/(\hbar \Omega_0)\gg 1$. The variances 
are nomalized by their asymptonic values, given by 
Eqs. (\ref{longtimephi}) and (\ref{longtimezeta}). 
The parameter $\gamma_Q$ is defined in Eq. (\ref{luca-gammaq}). 
{\cred The two upper solid and dashed curves correspond to $\gamma_Q=1/3$, 
the two intermediate ones to $\gamma_Q=1/5$, and the two lower ones 
to $\gamma_Q=1/10$.}} 
\label{fig3}
\end{figure}

\subsection{High temperature}

Let us start from the relative phase fluctuations at high temperature 
with Eqs. \eqref{hightempxiphicorr} and \eqref{hightempxizetacorr}. 
Eqs. \eqref{generalsolphi}, \eqref{hightempxiphicorr}, 
and \eqref{hightempxizetacorr} give
\beqa 
\Delta\phi_0(t)^2 &=& \Delta\phi_0(\infty)^2 \, 
\Bigg[ 1 + e^{-2\gamma_Q \Omega_0 t} 
\biggl[ \frac{\gamma_Q^2 - \gamma_J^2}{\gamma_J^2} 
\nonumber 
\\
&-& {\gamma_Q^2\over\gamma_J^2} 
\cos(2\gamma_J \Omega_0 t) - {\gamma_Q\over\gamma_J} \sin(2\gamma_J \Omega_0 t) 
 \biggl] \Bigg] ,
\label{hightempdeltaphi} 
\eeqa
\beqa
\Delta\zeta_0(t)^2 &=& \Delta\zeta_0(\infty)^2 
\Bigg[ 1 - 
e^{-2\gamma_Q \Omega_0 t}\biggl[ \frac{1}{\gamma_J^2} 
\nonumber 
\\
&-& {\gamma_Q^2\over\gamma_J^2} \cos(2\gamma_J \Omega_0 t) 
+ {\gamma_Q\over\gamma_J} \sin(2\gamma_J \Omega_0 t) \biggl] \Bigg] ,
\label{hightempdeltazeta}
\eeqa
where the asymptotic values of these variances
for $t\to +\infty$ are given by 
\beqa
\Delta\phi_0(\infty) &=& 
\frac{1}{\Omega_0}\sqrt{\frac{k_{\rm B}T}{2ML^2}} ,
\label{longtimephi} 
\\
\Delta\zeta_0(\infty) &=& 
\sqrt{\frac{Mk_{\rm B}T}{2\hbar^2\bar\rho^2}} .\label{longtimezeta}
\eeqa
Notice that Eq. (\ref{hightempdeltazeta}) is equivalent to the one 
in Ref. \cite{minguzzi}.

By using Eqs. (\ref{hightempdeltaphi}) and (\ref{hightempdeltazeta}) 
we plot the time evolution of the fluctuations 
in the relative phase and population imbalance in Fig. \ref{fig3}. 
They exhibit strong
similarities in the behaviour of these two quantities, which are ruled by
the damping parameter $\gamma_Q$. One can see that curves
tend to overlap as much as one lowers $\gamma_Q$.

\subsection{Generic temperature}

Let us focus on the generic temperature case to see effects of 
quantum fluctuations in addition to thermal ones.
We already mentioned the fact that the noise correlators in the most general 
case are complex valued quantities. Then we use
Eq. (\ref{xicomplexexpansion}) for both $\expval{\xi_\phi(t)\xi_\phi(t')}$
and $\expval{\xi_\zeta(t)\xi_\zeta(t')}$ and we obtain
\begin{align}
& \Delta\phi_0(t)^2 = \frac{1}{2}\int_{0}^{t}ds \int_{0}^{t}ds' 
\expval{\acomm{\xi_\phi(s)}{\xi_\phi(s')}} \chi(t-s)\chi(t-s') , 
\label{deltaphiintegral}\\
& \Delta\zeta_0(t)^2 = \frac{1}{2}\int_{0}^{t}ds\int_{0}^{t}ds' 
\expval{\acomm{\xi_\zeta(s)}{\xi_\zeta(s')}} \chi(t-s)\chi(t-s') . 
\label{deltazetaintegral}
\end{align}
Eq. (\ref{FTsymmxiphi}) provides
\beqa
\Delta\phi_0(t)^2 &=& \frac{\gamma_0}{M\Omega_0^2}\int_{-\infty}^{+\infty}
d\omega \hbar\omega^3 \coth(\frac{\beta\hbar\omega}{2})
\nonumber 
\\
&\times& \int_{0}^{t}dt_1 
\chi(t_1)e^{-i\omega t_1} \int_{0}^{t}dt_2 \chi(t_2)e^{i\omega t_2} ,
\label{firstintdeltaphi}
\eeqa
where we changed the time variables $t_1 = t-s$ and $t_2=t-s'$. 
Introducing a new adimensional frequency $\tilde\omega = \omega/\Omega_0$, 
we obtain an expression for the variance of the phase 
\beq
\Delta\phi_0(t)^2 = \Gamma \frac{\gamma_Q}{\gamma_J^2}
\int_{-\infty}^{+\infty}d\tilde\omega \tilde\omega^3 
\coth(\frac{\beta\hbar\Omega_0\tilde\omega}{2}) 
G_{\tilde\omega}(t)G_{-\tilde\omega}(t) ,
\label{lowtempvarphi}
\eeq
where 
\beq 
\Gamma = {2\hbar\over M\Omega_0} ,
\label{luca-Gamma}
\eeq
\begin{figure}[t]
\centerline{\epsfig{file=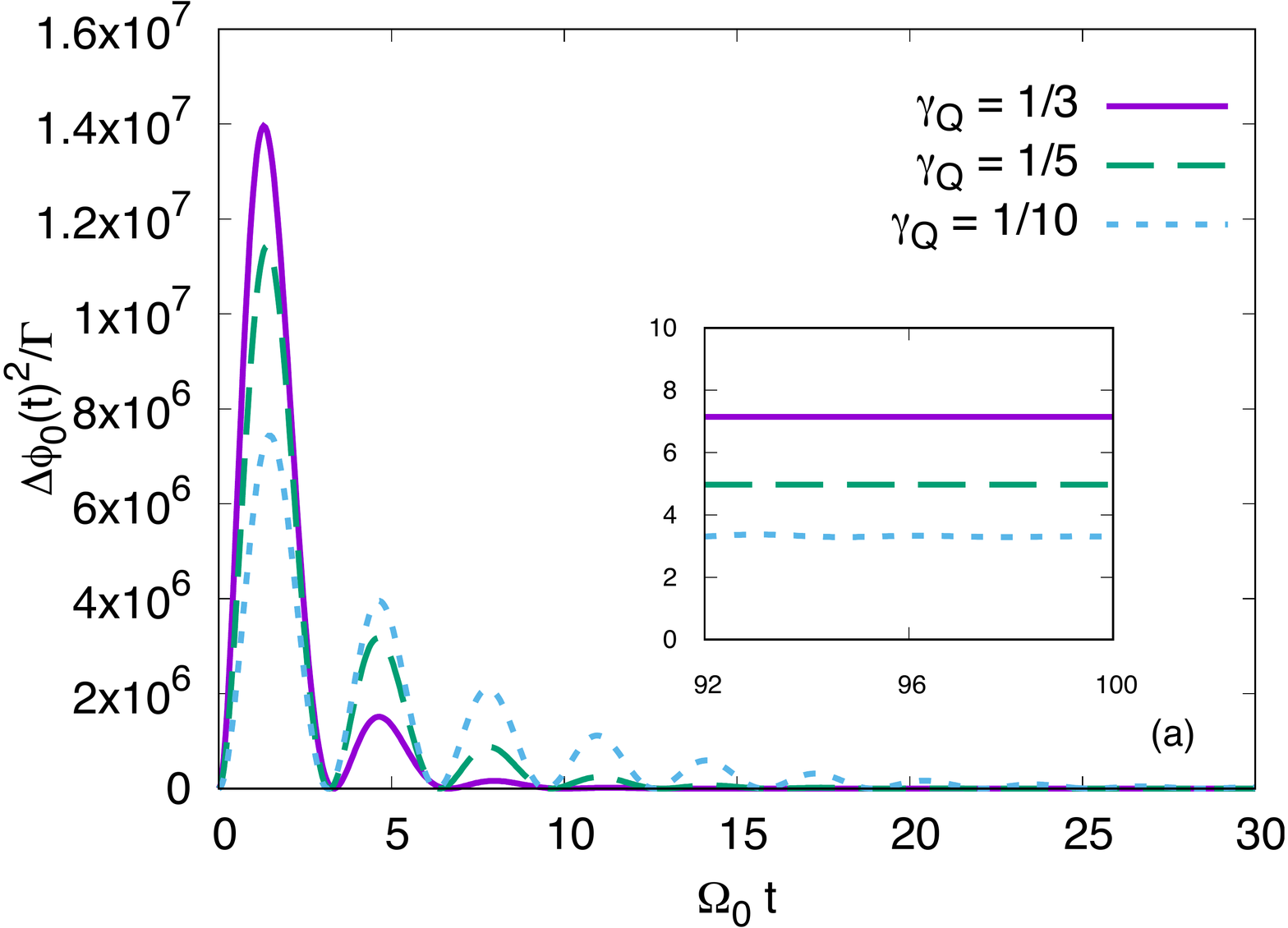,width=1.05\linewidth}}
\centerline{\epsfig{file=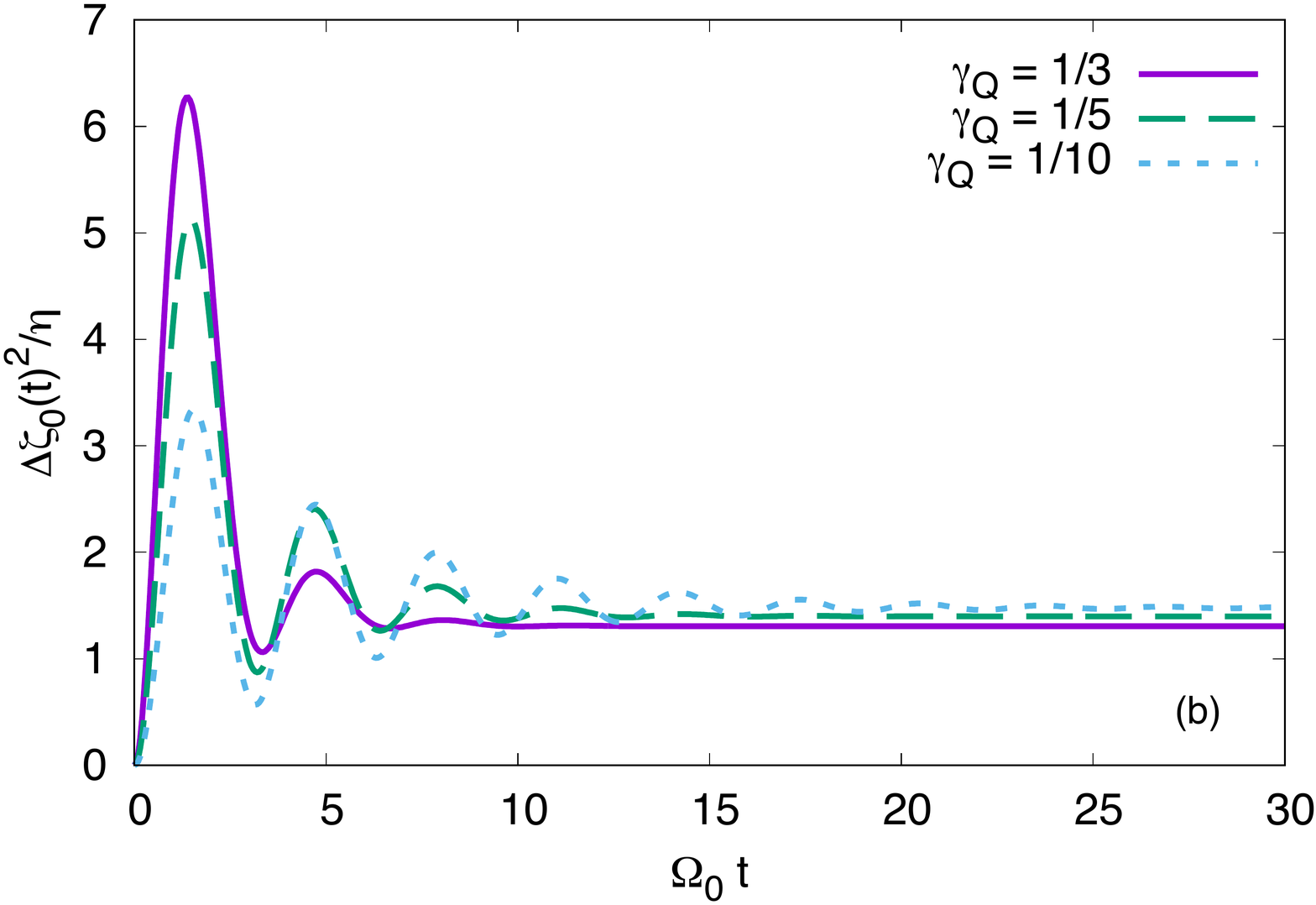,width=1\linewidth}}
\small 
\caption{(a) Time evolution of the relative phase variance 
$\Delta\phi_0(t)^2$ of the Josephson mode. (b) Time evolution of the population 
imbalance variance $\Delta\zeta_0(t)^2$ of the Josephson mode. 
Here we set three different values for $\gamma_Q$ 
and $k_{\rm B}T/(\hbar\Omega_0)=10^{-2}$ (low-temperature regime). 
The normalization factors $\Gamma$ and $\eta$ are defined 
in Eqs. (\ref{luca-Gamma}) and 
(\ref{luca-eta}). Note that the results are 
crucially dependent on an ultraviolet cutoff ${\tilde \omega}_{\rm max}$. 
Here we use ${\tilde \omega}_{\rm max}=10^4$ (see the main text for details).} 
\label{fig4}
\end{figure}
and 
\beqa
G_{\tilde\omega}(t) &=& \frac{e^{-(i\tilde\omega + \gamma_Q)\Omega_{0}t}}
{(i\tilde\omega + \gamma_Q)^2 - \gamma_J^2}
\Big[ \gamma_Je^{(i\tilde\omega + \gamma_Q)(\Omega_0t)} 
\nonumber 
\\
&-& \gamma_J\cos(\gamma_J \Omega_0t) 
- (i\tilde\omega + \gamma_Q)\sin(\gamma_J \Omega_0t) \Big] .
\eeqa
What we have just seen can be identically translated into the 
calculation of Eq. (\ref{deltazetaintegral}) as
\beq
\Delta\zeta_0(t)^2 = \eta \frac{\gamma_Q}{\gamma_J^2}
\int_{-\infty}^{+\infty}d\tilde\omega \tilde\omega 
\coth(\frac{\beta\hbar\Omega_0 \tilde\omega}{2}) 
G_{\tilde\omega}(t)G_{-\tilde\omega}(t) ,
\label{lowtempvarzeta}
\eeq
where 
\beq 
\eta = {M\Omega_0\over \hbar\bar\rho^2} .
\label{luca-eta}
\eeq
Both of Eqs. (\ref{lowtempvarphi}) and (\ref{lowtempvarzeta}) are 
computed numerically, 
but it is worth noting that they have an ultraviolet divergence. 
For this reason it is necessary 
to set a cutoff that, taking into account the phonon dispersion relation 
we found previously, can be seen as ${\tilde\omega}_{\rm max}= 
c_s\pi N/(\Omega_0 L)$. Using typical parameters like 
$N=10^3$, $L= 10^{-6}{\rm m}$, $\Omega_0= 10^2{\rm s^{-1}}$, 
$c_s= 10^{-3}{\rm ms^{-1}}$ \cite{pigneur}, we find 
${\tilde\omega}_{\rm max} = 10^4$.

In Fig. \ref{fig4}, we report the time evolution of the 
fluctuations for three values of $\gamma_Q$ in the underdamped 
regime ($\gamma_Q<1)$. Panel (a) in Fig. \ref{fig4} clearly shows that 
the variance $\Delta\phi_0(t)^2$, related 
to the relative phase of the Josephson mode, oscillates until it
reaches an asymptotic value in the long-time limit, as shown in the inset. 
The huge initial oscillations
are crucially dependent on the cutoff $\tilde\omega_{\rm max}$, and 
the asymptotic value strongly depends on the choice of $\gamma_Q$. 
It is worth noting that decreasing 
the value of $\gamma_Q$ leads to a lower asymptotic value. 
The variance $\Delta\zeta_0(t)^2$ 
of the population imbalance of the Josephson mode in panel (b) 
of Fig. \ref{fig4} also oscillates in time, reaching asymptotically 
a finite value. Here the asymptotic value also strongly depends 
on the damping parameter $\gamma_Q$ and decreasing $\gamma_Q$ leads to 
a larger asymptotic value.

\begin{figure}[t]
\centerline{\epsfig{file=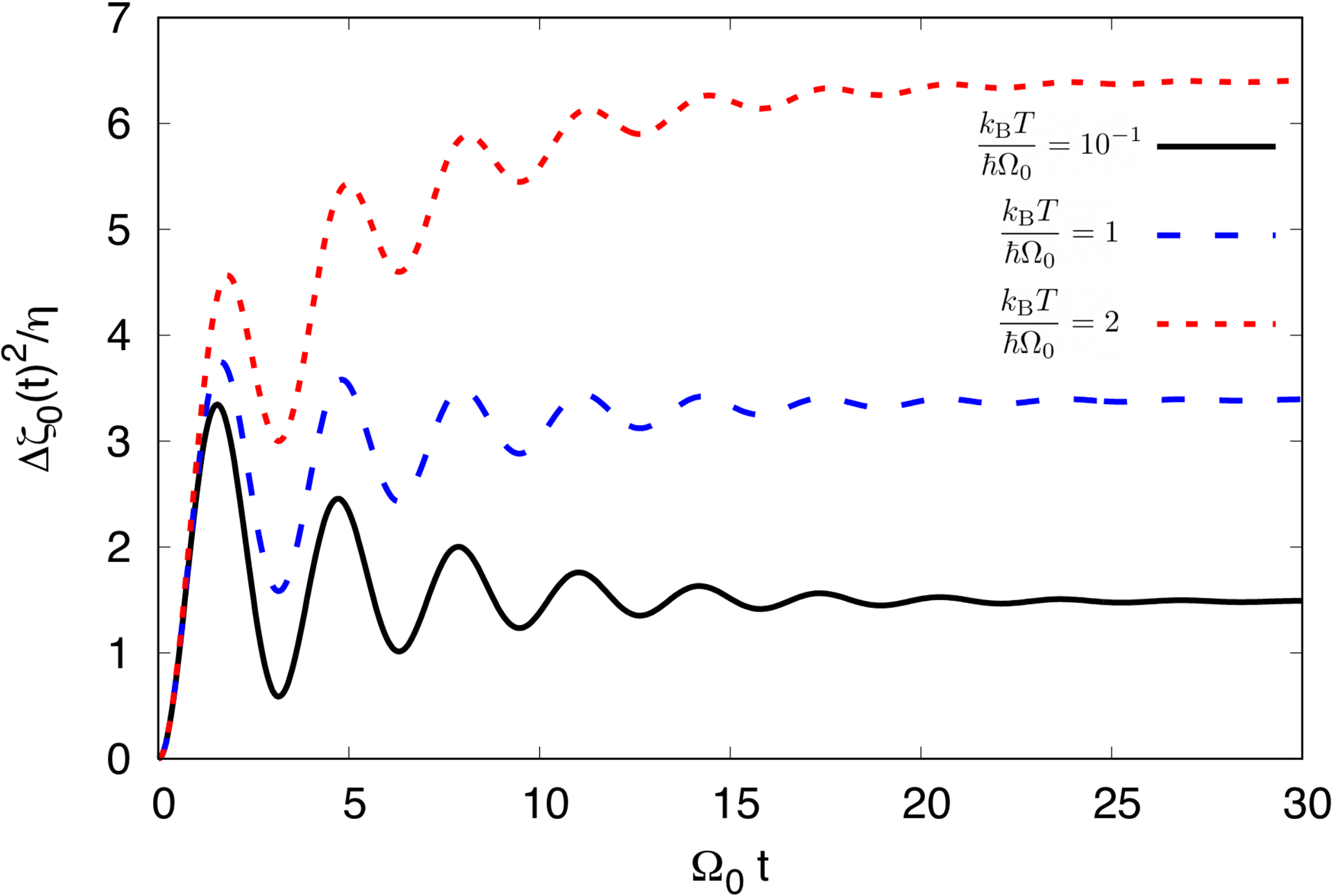,width=1.05\linewidth}}
\small 
\caption{Variance $\Delta\zeta_0(t)^2$ of the Josephson 
population imbalance as a function of time $t$ for three 
values of the temperature $T$ of the bath of phonons. 
The damping coefficient $\gamma_Q$ is set to be $\gamma_Q=1/10$. 
$\Omega_0$ is the Josephson frequency of Eq. (\ref{imp-fre}) 
while $\eta$ is given by Eq. (\ref{luca-eta}).}
\label{fig5}
\end{figure}

For the sake of completeness, in Fig. \ref{fig5}, we plot 
$\Delta\zeta_0(t)^2$ for three values of the rescaled 
temperature $k_{\rm B}T/(\hbar\Omega_0)$ of the phonon bath. 
The figure shows that the asymptotic value 
$\Delta\zeta_0(\infty)^2$ grows by increasing the temperature. 
We have verified that the same happens with $\Delta\phi_0(\infty)^2$. 
These results are fully consistent with the ones obtained 
in Ref. \cite{grabert} for the quadratic fluctuations of the damped 
harmonic oscillator in thermal equilibrium. 

\section{Conclusions}

We have found that in elongated Josephson junctions 
the damped dynamics of the Josephson mode 
is ruled by the damping constant $\gamma_0$, given by Eq. (\ref{imp-gamma}). 
Moreover, the Josephson dynamics is strictly dependent on the interaction 
between the Josephson mode and the quantum-thermal bath of phonons. 
We have studied the phase fluctuation 
$\Delta\phi_0(t)$ and the population-imbalance fluctuation 
$\Delta\zeta_0(t)$ of the Josephson mode. 
{\cred In our work, as well as in Ref. \cite{minguzzi}, 
both noise and damping have the same origin: 
the intrinsic coupling to the phonon bath. The damping of correlations 
is caused by the friction while 
nonzero variances are ascribed to the noise.} 
In the high-temperature regime, 
where the thermal fluctuations dominate, we have 
derived analytic expressions for the Josephson fluctuations. 
The generic temperature case has also been taken into account. 
Here quantum fluctuations play an important 
role and we rely on numerical calculations to determine 
the time evolution of the Josephson fluctuations, which exhibit 
their thermalization to constant values after a transient 
characterized by oscillating dynamics.

\section*{Acknowledgments} 

The authors acknowledge Anna Minguzzi and Juan Polo 
for useful e-clarifications. LS thanks Andrea Tononi and Flavio Toigo 
for enlightening discussions. KF is supported by a PhD fellowship 
of the Fondazione Cariparo. 

\appendix

\section{Derivation of Eq. \eqref{Qnt}}\label{AppA}

In this section, we briefly explain how to obtain the solution of the 
bath coordinates in Eq. \eqref{Qnt}.
Taking the derivative of the first two equations of motion in 
Eqs. \eqref{dotQ0} and \eqref{dotQn} with respect to time, we reach 
two differential equations for the Josephson mode and the excited modes as
\begin{align}
&\ddot Q_n(t) + \omega_n^2Q_n(t) = -\frac{J_0\bar\rho}{ML}
\sin(\frac{Q_0(t)}{L}) ,\label{Qndiff} \\
&\ddot Q_0(t) + \frac{J_0\bar\rho}{ML}\sin(\frac{Q_0(t)}{L}) = 
\sum_{n=1}^{+\infty}\ddot Q_n(t) .\label{Q0diff}
\end{align}
In this way we can start by solving the first equation, finding $Q_n(t)$, 
and then we exploit what we found in order to study the second one.
Eq. (\ref{Qndiff}) can be solved by taking the Laplace transformation 
on each member, and we end up with
\beq
\begin{split}
Q_n(t)& = \cos(\omega_n t)Q_n(0) + \frac{\sin(\omega_n t)}{\omega_n}
\dot Q_n(0) + \\
& - \frac{J_0\bar\rho}{ML\omega_n}\int_{0}^{t}dt' 
\sin{\left[\omega_n (t-t')\right]}
\sin(\frac{Q_0(t')}{L}) .
\end{split}
\eeq
Integrating by parts the last term, we obtain Eq. \eqref{Qnt}.


\begin{thebibliography}{999}

\bibitem{caldeira} A. O. Caldeira and A. J. Leggett, 
Phys. Rev. Lett. \textbf{46}, 211 (1981).

\bibitem{ford} G. W. Ford, J. T. Lewis, and R. F. O'Connell, 
Phys. Rev. A \textbf{37}, 4419 (1988).

\bibitem{schmidt} A. Schmidt, Phys. Rev. Lett. {\bf 51}, 1506 (1983).

\bibitem{bulgadaev} S. A. Bulgadaev, JETP Lett. {\bf 39}, 315 (1984).

\bibitem{dalla} E. D. Torre, E. Demler, T. Giamarchi, and E. Altman, 
Nat. Phys. {\bf 6}, 806 (2010).

\bibitem{koch80} R. H. Koch, D. J. Van Harlingen, and J. Clarke, 
Phys. Rev. Lett. {\bf 45}, 2132 (1980).

\bibitem{koch82} R. H. Koch, D. J. Van Harlingen, and J. Clarke, 
Phys. Rev. B {\bf 26}, 74 (1982).

\bibitem{pigneur} M. Pigneur, T. Berrada, M. Bonneau, T. Schumm, E. Demler, 
and J. Schmiedmayer, Phys. Rev. Lett. \textbf{120}, 173601 (2017).

\bibitem{minguzzi} J. Polo, V. Ahufinger, F. W. J. Hekking, and A. Minguzzi, 
Phys. Rev. Lett. \textbf{2}, 090404 (2018).

\bibitem{tononi} A. Tononi, F. Toigo, S. Wimberger, A. Cappellaro, 
and L. Salasnich, New J. Phys. {\bf 22}, 073020 (2020).

{\cred 
\bibitem{essler} Y. D van Nieuwkerk and F. H. L. Essler, 
SciPost Phys. {\bf 9}, 025 (2020). 

\bibitem{fendley} P. Fendley, F. Lesage, and H. Saleur, 
J. Stat. Phys. {\bf 85}, 211 (1996). 

\bibitem{bouchoule} I. Bouchoule, Eur. Phys. J. D {\bf 35}, 147 (2005).

\bibitem{grisins} P. Grisins and I. E. Mazets, 
Phys. Rev. A {\bf 87}, 013629 (2013).
}

\bibitem{javanainen} J. Javanainen, Phys. Rev. Lett. \textbf{57}, 3164 (1986). 

\bibitem{josephson1962} B. D. Josephson, Phys. Lett. \textbf{1}, 251 (1962).

\bibitem{smerzi} A. Smerzi, S. Fantoni, S. Giovanazzi, and S. R. Shenoy, 
Phys. Rev. Lett. \textbf{79}, 4950 (1997).

\bibitem{oberthaler} M. Albiez, R. Gati, J. Folling, S. Hunsmann, 
M. Cristiani, and M.K. Oberthaler, Phys. Rev. Lett. \textbf{95}, 010402 (2005).

\bibitem{gersch} H. A. Gersch and G. C. Knollman, 
Phys. Rev. \textbf{129}, 959 (1963).

\bibitem{ingold} G. Ingold, A. Buchleitner, and K. Hornberger, 
\textit{Coherent Evolution in Noisy Environments} (Springer, 2002). 

\bibitem{grabert} H. Grabert, U, Weiss, and P. Talkner, 
Z. Phys. B {\bf 55}, 87 (1984). 

\end{thebibliography}
\end{document}